\documentclass[a4paper,11pt]{article}
\pdfoutput=1 

\usepackage{jheppub} 
\usepackage[T1]{fontenc} 

\usepackage{graphicx}  
\usepackage{dcolumn}   
\usepackage{bm}        
\usepackage{amssymb}   
\usepackage{amsmath}

\usepackage{hyperref}

\usepackage{hhline}
\usepackage{xcolor}

\renewcommand\vec[1]{\ensuremath\boldsymbol{#1}} 

\title{Quantum Electrodynamics of Non-Hermitian Dirac Fermions}

\author[a]{Sk Asrap Murshed,}
\author[a,1]{Bitan Roy, \note{Corresponding author.}}

\affiliation[a]{Department of Physics, Lehigh University, Bethlehem, Pennsylvania, 18015, USA}

\date{\today}

\emailAdd{skm421@lehigh.edu}
\emailAdd{bitan.roy@lehigh.edu}

\abstract{We develop an effective quantum electrodynamics for non-Hermitian (NH) Dirac materials interacting with photons. These systems are described by nonspatial symmetry protected Lorentz invariant NH Dirac operators, featuring two velocity parameters $v_{_{\rm H}}$ and $v_{_{\rm NH}}$ associated with the standard Hermitian and a masslike anti-Hermitian Dirac operators, respectively. They display linear energy-momentum relation, however, in terms of an effective Fermi velocity $v_{_{\rm F}}=\sqrt{v^2_{_{\rm H}}-v^2_{_{\rm NH}}}$ of NH Dirac fermions. Interaction with the fluctuating electromagnetic radiation then gives birth to an emergent Lorentz symmetry in this family of NH Dirac materials in the deep infrared regime, where the system possesses a unique terminal velocity $v_{_{\rm F}}=c$, with $c$ being the speed of light. While in two dimensions such a terminal velocity is set by the speed of light in the free space, dynamic screening in three spatial dimensions permits its nonuniversal values. Manifestations of such an emergent spacetime symmetry on the scale dependence of various physical observables in correlated NH Dirac materials are discussed.}


\arxivnumber{2309.07916}

\begin{document}

\maketitle
\flushbottom

\section{Introduction}

Quantum electrodynamics (QED) is one of the most successful theoretical description of the light-matter interaction~\cite{Peskin2019}. Among its many triumphs, perhaps the most prominent one is the highly accurate prediction of the radiative correction for the $g$ factor of electrons due the interaction with photons~\cite{Schwinger1948}. Even the leading-order (one-loop) correction agrees remarkably well with experiments. What plays an important role in the formulation of QED and the subsequent development of the electroweak and strong forces, ultimately leading to the Standard model of fundamental particles, is the spacetime Lorentz symmetry. It ensures a single velocity for all the participating degrees of freedom, the speed of light ($c$), realized at sufficiently high energies~\cite{Nielsen1978, Kostelecky2011}. Dynamics of the underlying fermionic degrees of freedom in the Standard model is governed by the Dirac equation~\cite{Dirac1928}, manifesting a linear energy-momentum relation, and thus a Lorentz invariance.

Often crystalline quantum materials also foster linearly dispersing emergent fermionic quasiparticles at low energies around a few isolated (and special) points in the Brillouin zone~\cite{Castroneto2009, Balatsky2014, Armitage2018}. The Fermi velocity ($v_{_{\rm F}}$), typically much smaller than $c$, of such low energy quasiparticles in this family of materials plays the role of the speed of light, with $v_{_{\rm F}}=c/300$ in graphene, for example. On the other hand, when Dirac quasiparticles interact with photons, ultimately the system arrives at an infrared stable renormalzation group (RG) fixed point where $v_{_{\rm F}}=c$, thereby showcasing an emergent spacetime Lorentz symmetry in the crystalline universe, although it is generically absent at the lattice or ultraviolet scale~\cite{Chadha1983, Gonzalez1994, Isobe2012, RoyJHEP2016}.

While these outcomes are extremely robust and well appreciated in isolated or closed Dirac crystals, the fate of QED and the possible emergence of the Lorentz symmetry in the presence of system-to-environment interactions in non-Hermitian (NH) Dirac materials remain unanswered so far. We here address this issue by considering a NH Dirac operator, constructed by supplementing the standard Hermitian Dirac kinetic term, appearing with a velocity parameter $v_{_{\rm H}}$, by a masslike anti-Hermitian Dirac kinetic operator with another velocity parameter $v_{_{\rm NH}}$, such that $v_{_{\rm H}}>v_{_{\rm NH}}$. The later one is realized by taking a product between mutually anticommuting Hermitian mass matrix ($M$) and the Dirac kinetic operator. See Eq.~\eqref{eq:NHDirac}. Throughout, we assume that the effective Fermi velocity of a collection of such linearly dispersing NH Dirac fermions ($v_{_{\rm F}}=\sqrt{v^2_{_{\rm H}}-v^2_{_{\rm NH}}}$) and the speed of light ($c$) are different at the microscopic level, and $v_{_{\rm F}}<c$. Then, from the leading-order perturbative analysis of the NH Dirac fermions interacting with photons [see Fig.~\ref{fig:Self-energy}], here we show that both in two and three spatial dimensions $v_{_{\rm F}}$ always approaches the speed of light in the deep infrared regime through radiative quantum corrections. See Figs.~\ref{fig:RGflow2D} and~\ref{fig:RGflow3D}, detailed in Appendix~\ref{SMSec:FermionicSE}, Appendix~\ref{SMSec:PhotonSE} and Appendix~\ref{SMSec:Vertex}. In two spatial dimensions ($d=2$), while the speed to light does not renormalize, in $d=3$ it gets reduced under coarse grain due to the many-body dynamic screening. As a consequence, the unique terminal velocity in NH Dirac materials is \emph{nonuniversal} in $d=3$. Even though a one-to-one correspondence between the nature of the system-to-environment interaction, giving rise to dissipation (source of non-Hermiticity in any open quantum system) and the resulting NH operator is missing, as the same NH operator may result from many different realizations of system-to-environment interaction, we arrive at these conclusions by considering symmetry protected Lorentz invariant NH Dirac operators in $d=2$ and $d=3$. These outcomes may thus suggest a possible generic emergence of the Lorentz symmetry in open NH Dirac crystals. We also outline the impact of such an emergent spacetime symmetry on the scaling behavior of various experimentally measurable physical observables. Furthermore, in Appendix~\ref{SMSec:NHgraphene}, we outline a simple possible realization of NH Dirac fermions in graphene, a key member of the planar Dirac material family, resulting from a unidirectional hopping imbalance between two sublattices of the honeycomb lattice. This construction can in principle be realized on optical honeycomb lattices of neutral atoms~\cite{Esslinger2013}, in the presence of dissipation due to a coupling with the environment or bath, by generalizing a proposed construction to a similar situation on one-dimensional chain from Appendix~F of Ref.~\cite{GongPRX2018}, which we also discuss in Sec.~\ref{sec:summary}. However, experimental demonstration of QED in NH Dirac materials demands a similar construction in electronic systems, for which designer electronic materials constitute the most promising platform, on which graphene has already been synthesized~\cite{Manoharan2012}.

\subsection{Organization}

The rest of the paper is organized as follows. In the next section (Sec.~\ref{sec:NHDirac}), we introduce non-spatial symmetry protected Lorentz invariant NH Dirac operator in any dimension. Sec.~\ref{sec:NHQED} is devoted to the quantum electrodynamics in NH Dirac materials and the corresponding leading-order or one-loop Feynman diagrams. Emergent Lorentz symmetry resulting from the interactions between NH Dirac fermions and photons in two and three dimensions are shown in Sec.~\ref{sec:Lorentz2D} and Sec.~\ref{sec:Lorentz3D}, respectively. We summarize the findings and present discussions on related issues in Sec.~\ref{sec:summary}. Explicit computations of the Feynman diagrams, shown in Fig.~\ref{fig:Self-energy}, are displayed in first three appendices, and a microscopic realization of NH Dirac fermions in graphene is presented in Appendix~\ref{SMSec:NHgraphene}.

\section{Non-Hermitian Dirac operator}~\label{sec:NHDirac}

The Lorentz invariant NH Dirac operator in $d$ spatial dimensions is given by~\cite{RoyJuricicNHDSM2023} 
\begin{equation}~\label{eq:NHDirac}
H_{\rm NH} (\vec{k})=v_{_{\rm H}} \left( \sum^d_{j=1} \Gamma_j k_j \right) + v_{_{\rm NH}}\; \left( M \sum^d_{j=1} \Gamma_j k_j \right), 
\end{equation}
where $\vec{k}=(k_1, \cdots, k_d)$ is the momentum, and $v_{_{\rm H}}$ and $v_{_{\rm NH}}$ bear the dimension of the Fermi velocity. The internal structure of the associated Dirac spinor ($\Psi$) depends on the microscopic details, which we do not delve into here. For concreteness, we work with the minimal four component Dirac system. The Hermitian matrices satisfy the anticommuting Clifford algebra $\{ \Gamma_j, \Gamma_k \}=2\delta_{jk}$ and $\{\Gamma_j, M \}=0$ for $j,k=1, \cdots, d$ with $M^2=1$. The first term in $H_{\rm NH}(\vec{k})$ corresponds to the standard Dirac Hamiltonian yielding the conventional linear real energy-momentum relation $E_{\rm H}=\pm v_{_{\rm H}} |\vec{k}|$. The second term in $H_{\rm NH}(\vec{k})$, on the other hand, is anti-Hermitian by construction as $(M \Gamma_j)^\dagger=-M \Gamma_j$ for any $j$. These two terms anticommute with each other. Therefore, the anti-Hermitian operator acts as a momentum-dependent masslike term to regular Dirac fermions. As a consequence, the eigenvalue spectrum of $H_{\rm NH}$ is given by $E_{\rm NH}=\pm \sqrt{v^2_{_{\rm H}}-v^2_{_{\rm NH}}} |\vec{k}| \equiv v_{_{\rm F}} |\vec{k}|$, where $v_{_{\rm F}}=\sqrt{v^2_{_{\rm H}}-v^2_{_{\rm NH}}}$ is the effective Fermi velocity of NH Dirac fermions. The eigenvalues of $H_{\rm NH}(\vec{k})$ are, therefore, purely real and purely imaginary for $v_{_{\rm H}}> v_{_{\rm NH}}$ and $v_{_{\rm H}}< v_{_{\rm NH}}$, respectively. Throughout, we assume the former situation. The mass matrix $M$ fully anticommutes with the standard Dirac Hamiltonian and its NH counterpart. Both the standard Dirac kinetic term [first quantity in Eq.~\eqref{eq:NHDirac}] and the Dirac mass operator ($M$) are Hermitian Lorentz scalars. As a consequence, their product [second quantity in Eq.~\eqref{eq:NHDirac}] is also Lorentz invariant, but anti-Hermitian as these two Hermitian operators mutually anticommute. In Appendix~\ref{SMSec:NHgraphene}, we show one microscopic realization of such NH Dirac operator in graphene's honeycomb lattice for a specific choice of $M$. A uniform expectation value of such a mass operator in the ground state, i.e.\ when $\langle \Psi^\dagger M \Psi \rangle \neq 0$, gives rise to an isotropic gapped state. However, in NH Dirac systems condensation of any mass order must always break at least one of its four \emph{nonspatial}, namely time-reversal, anti-unitary and unitary particle-hole, and pseudo-Hermiticity symmetries~\cite{Bernard2002}. Thus in any dimensions, the nodal NH Dirac quasiparticles are symmetry protected and cannot be gapped unless at least one of these four nonspatial symmetries is broken~\cite{RoyJuricicNHDSM2023}. Next, we consider interaction between such gapless system and photons, leading to QED for NH Dirac materials.

\begin{figure}[t!]
\includegraphics[width=1.00\textwidth]{FeynmanDiagrams.pdf}
\caption{Leading-order (one-loop) self-energy diagrams for (a) non-Hermitian Dirac fermions (solid lines) and (b) photons (wavy lines), and the one-loop vertex correction is shown in (c). The explicit form of their interaction vertex (indexed by $\rho$, $\lambda$ and $\mu$) is shown in Eq.~\eqref{eq:QEDvertex}.
}~\label{fig:Self-energy}
\end{figure}

\section{Quantum electrodynamic and Feynman diagrams}~\label{sec:NHQED}

The imaginary time ($\tau$) action describing the interaction between fluctuating electromagnetic fields and NH Dirac fermions takes the form 
\begin{equation}
S^{\rm QED}_{\rm NH}= \int d\tau d^D\vec{x} \; \left( L_{\rm F} + L_{\rm EM} \right),
\end{equation}
where $D=d-\epsilon$ about which more in a moment, 
\begin{equation}
L_{\rm F} = \Psi^\dagger \left[ \left( \partial_0 -i e A_0 \right) + H_{\rm NH} \left( \vec{k} \to -i {\boldsymbol \nabla}-\frac{e}{c} \vec{A} \right) \right] \Psi,
\end{equation}
and $e$ is the electronic charge. The Maxwellian Lagrangian in three spatial dimensions is~\cite{Jackson1999} 
\begin{equation}
L_{\rm EM}=\frac{1}{4} F_{\rho \lambda} F_{\rho \lambda},
\end{equation}
where $\rho, \lambda=0, \cdots, d$ and $F_{\rho \lambda}=\partial_\rho A_\lambda -\partial_\lambda A_\rho$ is the electromagnetic field strength tensor. Here, $\Psi^\dagger \equiv \Psi^\dagger(\tau,\vec{x})$ and $\Psi \equiv \Psi(\tau,\vec{x})$ are two independent Grassmann variables, and $A_\rho \equiv A_\rho(\tau,\vec{x})$. The dynamics of NH Dirac fermions is set by the fermionic Green's function 
\begin{equation}~\label{eq:FemrionGreen}
G_{\rm F}(i\omega, \vec{k})=\frac{i \omega -H_{\rm NH}(\vec{k})}{\omega^2 + v^2_{_{\rm F}} |\vec{k}|^2}.
\end{equation}
The gauge field propagator in $d$ dimensions is given by 
\begin{equation}~\label{eq:PhotonGreen}
D_{\rho \lambda} (i\omega, \vec{k},d)= \delta_{\rho \lambda} \; \bigg(\varepsilon \left[\omega^2 + c^2  |\vec{k}|^2 \right]^{\frac{d-1}{2}}\bigg)^{-1},
\end{equation}
where the speed of light in the medium is $c=1/\sqrt{\varepsilon \mu}$, and $\varepsilon$ and $\mu$ are the permittivity and permeability of the medium, respectively. Here, $\omega$ is the Matsubara frequency. The non-analytic structure of the gauge field propagator in $d=2$ is obtained upon integrating the analytic one in $d=3$, which stems from the Maxwellian Lagrangian $L_{\rm EM}$~\cite{Jackson1999}, over the momentum in the $z$ direction ($k_z$) as follows 
\begin{eqnarray}
D_{\rho \lambda} (i\omega, \vec{k},2)
&=&\int^{\infty}_{-\infty}\frac{d (c k_z)}{2\pi} \: D_{\rho \lambda} (i\omega, \vec{k},3)
=\int^{\infty}_{-\infty}\frac{d (c k_z)}{2\pi} \: \frac{\delta_{\rho \lambda}}{\varepsilon \left[\omega^2 + c^2 \left(k^2_z + |\vec{k}_\perp|^2 \right) \right]} \nonumber \\
&=&\frac{\delta_{\rho \lambda}}{2 \varepsilon} \: \frac{1}{\left[\omega^2 + c^2 |\vec{k}_\perp|^2 \right]^{1/2}}
\end{eqnarray}
where $\vec{k}_\perp=(k_x,k_y)$ and we absorb the factor of 2 appearing in the denominator in $\varepsilon$, while defining $D_{\rho \lambda} (i\omega, \vec{k},2)$. Therefore, the speed of light does not renormalize in $d=2$, while it receives perturbative corrections due to many-body dynamic screening in $d=3$. These outcomes directly follow from the fact that the polarization bubble in $d=3$ and $d=2$ are respectively logarithmically divergent and finite, as shown in Appendix~\ref{SMSec:PhotonSE}. The interaction vertex between NH Dirac fermions and photons is given by $-i e \gamma_\rho$, where $\rho=0,\cdots,d$ and 
\begin{equation}~\label{eq:QEDvertex}
\gamma_\rho=\left(1, -\frac{i}{c} \left[ v_{_{\rm H}} \Gamma_j + v_{_{\rm NH}} M \Gamma_j \right] \right).
\end{equation}  
In what follows we compute the perturbative corrections to the velocity parameters $v_{_{\rm H}}$, $v_{_{\rm NH}}$ and $c$, arising from such QED interactions. To this end, we compute the self-energy diagrams for the NH Dirac fermions and photons, shown in Fig.~\ref{fig:Self-energy}, in $D$ spatial dimensions within the framework of an $\epsilon$ expansion, where $\epsilon=d-D$.

The contribution of the fermionic self-energy diagram, shown in Fig.~\ref{fig:Self-energy}(a), reads as 
\begin{eqnarray}
\Sigma(i\nu, \vec{k}) = (-i e)^2 \int \frac{d^D\vec{p}}{(2\pi)^D} \int^{\infty}_{-\infty} \frac{d\omega}{2 \pi} \gamma_\rho G_{\rm F} (i \omega+i \nu, \vec{p}+\vec{k}) \gamma_\lambda D_{\rho \lambda}(i \omega, \vec{p},d). 
\end{eqnarray}
The self-energy correction of the photon field (polarization bubble), shown in Fig.~\ref{fig:Self-energy}(b), yields 
\begin{eqnarray}
\Pi_{\rho \lambda} (i \nu,\vec{k}) = -(-i e)^2 \; {\rm Tr} \bigg[ \int \frac{d^D\vec{p}}{(2\pi)^D} \int^{\infty}_{-\infty} \frac{d\omega}{2 \pi} \gamma_\rho G_{\rm F} (i \omega, \vec{p}) \gamma_\lambda G_{\rm F} (i \omega+i \nu, \vec{p}+\vec{k}) \bigg].
\end{eqnarray}
Due to the distinct structure of the photon propagators in $d=2$ and $d=3$, we discuss these two cases separately. We compute the above integrals using the standard Feynman parametrization and identify their logarithmic divergence from $1/\epsilon$ poles~\cite{Peskin2019}. The computation of the fermionic self-energy is shown in Appendix~\ref{SMSec:FermionicSE}, while that for the polarization bubble is shown in Appendix~\ref{SMSec:PhotonSE}. The vertex diagram [Fig.~\ref{fig:Self-energy}(c)] does not lead to any new divergence, as shown in Appendix~\ref{SMSec:Vertex}. Its divergent piece is related to that stemming from the fermionic self-energy diagram, in agreement with the Ward identity~\cite{ward1950}, which so far was known to be applicable for Hermitian systems.

\begin{figure*}[t!]
\includegraphics[width=1.00\textwidth]{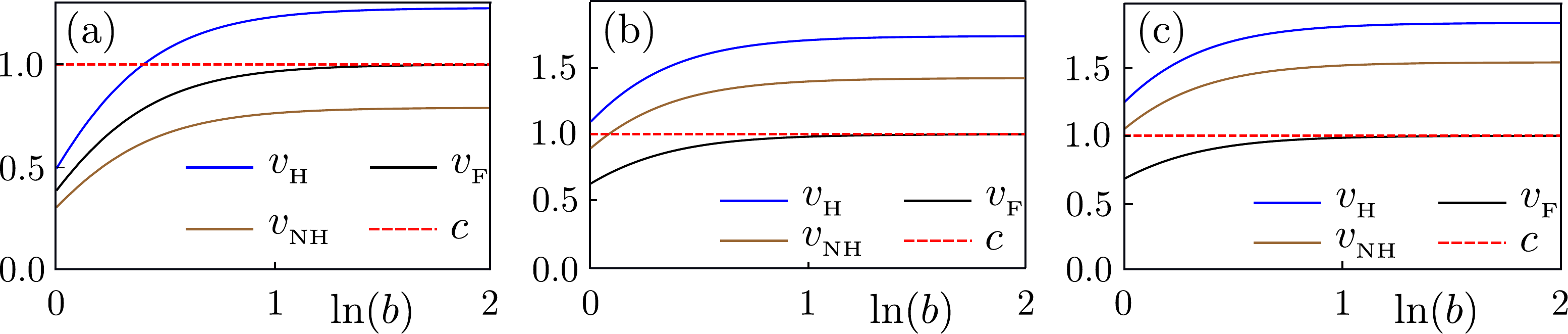}
\caption{Renormalization group (RG) flows of two velocity components of non-Hermitian Dirac fermions $v_{_{\rm H}}$ and $v_{_{\rm NH}}$ [Eq.~\eqref{eq:NHDirac}], and the effective Fermi velocity $v_{_{\rm F}}=\sqrt{v^2_{_{\rm H}}-v^2_{_{\rm NH}}}$ due to their interaction with photons in two dimensions for (a) $v^0_{_{\rm H}}=0.5$ and $v^0_{_{\rm NH}}=0.31$, such that $v^0_{_{\rm NH}}<v^0_{_{\rm H}}<c$, (b) $v^0_{_{\rm H}}=1.1$ and $v^0_{_{\rm NH}}=0.9$, such that $v^0_{_{\rm NH}}<c<v^0_{_{\rm H}}$, and (c) $v^0_{_{\rm H}}=1.25$ and $v^0_{_{\rm NH}}=1.05$, such that $c<v^0_{_{\rm NH}}<v^0_{_{\rm H}}$. Notice that $v^0_{_{\rm F}}< c$ (speed of light) always. The results are obtained by numerically solving Eq.~\eqref{eq:beta2D}. Throughout we measure all the velocities in units of $c$ (unrenormalized), which we set to be unity. Here, $b$ is the RG time. The quantities with the superscript `0' correspond to their bare values. For convenience, we set the fine structure constant $\alpha^{\rm 2D}_{_{\rm F}}=1$.     
}~\label{fig:RGflow2D}
\end{figure*}

\section{Non-Hermitian Lorentz symmetry in two dimensions}~\label{sec:Lorentz2D}

In two spatial dimensions the contribution from the polarization bubble is \emph{finite}, and thus the field renormalization condition for the gauge field ($A_\rho$) is trivial $Z_{A}=1$. By contrast, the fermionic self-energy diagram is logarithmically divergent, yielding the fermionic field renormalization 
\begin{equation}
Z_\Psi=1 + 2 \; \alpha^{\rm 2D}_{_{\rm F}} \; \frac{1- 2 x^2}{1-x^2} \; \left[ x-\frac{x^2 \; {\rm{cos^{-1}}}(x)}{(1-x^2)^{1/2}} \right] \frac{1}{\epsilon},
\end{equation}
where $x=v_{_{\rm F}}/c$ and $\alpha^{\rm 2D}_{_{\rm F}}=e^2/(4 \pi^2 \varepsilon v_{_{\rm F}}c)$ is the fine structure constant in two dimensions. The renormalization conditions for two velocity parameters of NH Dirac fermions are 
\begin{equation}
Z_Q=Z^{-1}_\Psi \left\{ 1 + \frac{\alpha^{\rm 2D}_{_{\rm F}}}{1-x^2} \bigg[x-\frac{{\rm{cos^{-1}}}(x)}{(1-x^2)^{1/2}} \bigg] \frac{1}{\epsilon} \right\}
\end{equation} 
for $Q=v_{_{\rm H}}$ and $v_{_{\rm NH}}$. Fascinatingly, both the velocity components receive identical perturbative correction, which is expressed solely in terms of the effective Fermi velocity ($v_{_{\rm F}}$) for NH Dirac fermions. Consequently, their RG flow equations or $\beta$ functions are given by  
\begin{equation}~\label{eq:beta2D}
\beta_Q = \alpha^{\rm 2D}_{_{\rm F}} \left[ \frac{x-4 x^3}{1-x^2} + \frac{1-2 x^2+4 x^4}{(1-x^2)^{3/2}} \; {\rm{cos^{-1}}}(x) \right] Q
\end{equation}
for $Q=v_{_{\rm H}}$ and $v_{_{\rm NH}}$, where $\beta_Q\equiv dQ/d\ln b$ and $b$ is the RG time. Remarkably, in Hermitian Dirac systems with $v_{_{\rm NH}}=0$ from the outset, the RG flow equation for the Fermi velocity $v_{_{\rm H}}=v_{_{\rm F}}$ takes the identical form~\cite{Gonzalez1994}. Such a similarity stems from the Lorentz invariant form of the NH Dirac operator [Eq.~\eqref{eq:NHDirac}]. The RG fixed point of the above flow equation for any arbitrary initial condition or bare value of $v_{_{\rm H}}$ and $v_{_{\rm NH}}$ is placed at $x=1$ or equivalently at $v_{_{\rm F}}=c$. As the speed of light does not renormalize in $d=2$, the terminal effective Fermi velocity for two-dimensional NH Dirac fermions in the deep infrared regime is always the speed of light in vacuum, with both the velocity components $v_{_{\rm H}}$ and $v_{_{\rm NH}}$ being nonzero, irrespective of their bare values, yielding NH Lorentz symmetry. We also confirm this outcome by numerically solving the RG flow equation from Eq.~\eqref{eq:beta2D}. The results are shown in Fig.~\ref{fig:RGflow2D}. Notice that in NH Dirac materials, the bare or renormalized values of $v_{_{\rm H}}$ and $v_{_{\rm NH}}$ can in principle be larger than $c$, while maintaining $v_{_{\rm F}} \leq c$ always, as long as the bare value of $v_{_{\rm F}}<c$. These scenarios are also shown in Fig.~\ref{fig:RGflow2D}. Therefore, two-dimensional massless NH Dirac quasiparticles always feature an emergent spacetime Lorentz symmetry with their terminal effective Fermi velocity being equal to the speed of light in the free space. We note that the velocity parameters associated with the Hermitian ($v_{_{\rm H}}$) and anti-Hermitian ($v_{_{\rm NH}}$) components of the the NH Dirac operator from Eq.~\eqref{eq:NHDirac}, respectively quantify nearest-neighbor tight-binding hopping amplitude for example on graphene's honeycomb lattice and the strength of dissipation or system-to-environment coupling, resulting for instance in a unidirectional hopping imbalance between the nearest-neighbor sites of graphene (see Appendix~\ref{SMSec:NHgraphene}). However, all the thermodynamic (density of states and specific heat), transport (optical conductivity) and elastic (optical shear viscosity) responses are solely determined by $v_{_{\rm F}}$~\cite{RoyJuricicNHDSM2023}, as discussed in Sec.~\ref{sec:summary}. Therefore, even when $v_{_{\rm H}}$ and $v_{_{\rm NH}}$ become \emph{superluminal} or larger than $c$ (possible, at least, in principle), causality is not violated since $v_{_{\rm F}}<c$ always. Similar outcomes hold in three dimensions, which we discuss next.

\section{Non-Hermitian Lorentz symmetry in three dimensions}~\label{sec:Lorentz3D}

In three dimensional NH Dirac systems the speed of light gets renormalized due to the interaction between the fermionic degrees of freedom and photons, causing dynamic many-body screening. Consequently, the polarization bubble [Fig.~\ref{fig:Self-energy}(b)] shows a logarithmic or an $1/\epsilon$ divergence, resulting in the nontrivial renormalization conditions for the permittivity and permeability of the medium, respectively give by
\begin{equation}~\label{eq:RGepsilonmu}
Z_{\varepsilon}= 1+ N_f \frac{4}{3} \frac{\alpha^{\rm 3D}_{_{\rm F}}}{\epsilon}
\;\; \text{and} \;\;
Z_{\mu}=1 - N_f \frac{4}{3} \frac{\alpha^{\rm 3D}_{_{\rm F}}}{\epsilon} x^2,
\end{equation}    
where $\alpha^{\rm 3D}_{_{\rm F}}=e^2/(8 \pi^2 \varepsilon v_{_{\rm F}} c^2)$ is the fine structure constant in three dimensions and $N_f$ is the number of four-component fermionic flavor. As the speed of light is related to these two quantities according to $c=1/\sqrt{\varepsilon \mu}$, we then arrive at the following RG flow equation of $c$
\begin{eqnarray}~\label{eq:Flowc3D}
\beta_c = - \frac{2}{3} N_f \alpha^{\rm 3D}_{_{\rm F}} \left( 1-x^2 \right) c.
\end{eqnarray}
The fine structure constant decreases monotonically under coarse grain according to
\begin{equation}~\label{eq:Flowalpha3D}
\beta_{\alpha^{\rm 3D}_{_{\rm F}}}=-\frac{4}{3} N_f \left( \alpha^{\rm 3D}_{_{\rm F}} \right)^2.
\end{equation} 
The RG flow equation for $c$ in a conventional Dirac system with $v_{_{\rm NH}}=0$ is identical to the above one, but in terms of the ratio $v_{_{\rm H}}/c$, as $v_{_{\rm F}}=v_{_{\rm H}}$ therein from the outset.

\begin{figure*}[t!]
\includegraphics[width=1.00\textwidth]{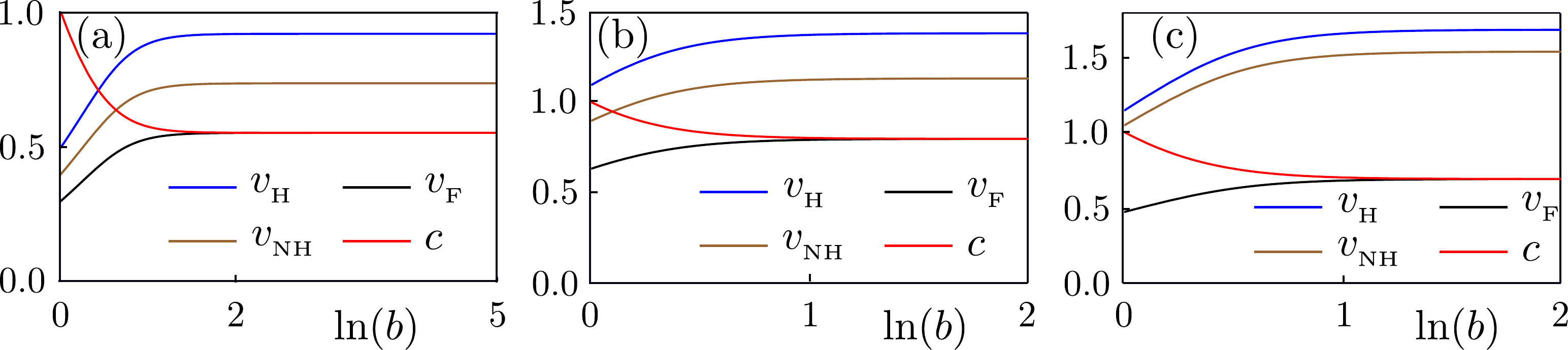}
\caption{Renormalization group (RG) flows of two velocity components of non-Hermitian Dirac fermions $v_{_{\rm H}}$ and $v_{_{\rm NH}}$ [Eq.~\eqref{eq:NHDirac}], as well as the effective Fermi velocity $v_{_{\rm F}}=\sqrt{v^2_{_{\rm H}}-v^2_{_{\rm NH}}}$ along with the speed of light in the medium ($c$) due to the gauge-fermion interaction in three dimensions for (a) $v^0_{_{\rm H}}=0.5$ and $v^0_{_{\rm NH}}=0.4$, such that $v^0_{_{\rm NH}}<v^0_{_{\rm H}}<c^0$, (b) $v^0_{_{\rm H}}=1.1$ and $v^0_{_{\rm NH}}=0.9$, such that $v^0_{_{\rm NH}}<c^0<v^0_{_{\rm H}}$, and (c) $v^0_{_{\rm H}}=1.15$ and $v^0_{_{\rm NH}}=1.05$, such that $c^0<v^0_{_{\rm NH}}<v^0_{_{\rm H}}$. Notice that $v^0_{_{\rm F}}< c$ always. The results are obtained by numerically solving Eqs.~\eqref{eq:Flowc3D},~\eqref{eq:Flowalpha3D} and~\eqref{eq:FlowVF3D}. Here, $b$ is the RG time and the quantities with the superscript `0' correspond to their bare values. Throughout we measure all velocities in units of $c^0$, which we set to be unity. For convenience, we set the bare fine structure constant $\alpha^{\rm 3D, 0}_{_{\rm F}}=1$ and $N_f=1$.
}~\label{fig:RGflow3D}
\end{figure*}

The fermionic field renormalization in $d=3$ is 
\begin{equation}
Z_\Psi=1-2 \; \alpha^{\rm 3D}_{_{\rm F}} \; \frac{x(1-3 x^2)}{(1+x)^2} \; \frac{1}{\epsilon}. 
\end{equation}
The renormalization conditions for two velocity parameters for three-dimensional NH Dirac fermions are 
\begin{eqnarray}
Z_Q= Z^{-1}_\Psi \left\{1-\frac{2}{3} \alpha^{\rm 3D}_{_{\rm F}} \; \frac{(1+x^2)(2+x)}{(1+x)^2} \; \frac{1}{\epsilon} \right\}, 
\end{eqnarray} 
for $Q=v_{_{\rm H}}$ and $v_{_{\rm NH}}$. Similar to the situation in $d=2$, the perturbative corrections for both the velocity components of NH Dirac fermions are identical in $d=3$ as well. Therefore, their RG flow equations or $\beta$ functions take identical form, given by 
\begin{equation}~\label{eq:FlowVF3D}
\beta_Q=\frac{4}{3} \frac{\alpha^{\rm 3D}_{_{\rm F}}}{(1+x)^2} \left[ 1 + 2 x +x^2-4 x^3 \right] \; Q,
\end{equation}
for $Q=v_{_{\rm H}}$ and $v_{_{\rm NH}}$. Notice that the pertubative correction in the above flow equations for $v_{_{\rm H}}$ and $v_{_{\rm NH}}$ is \emph{identical} to that for $v_{_{\rm H}}$ in conventional Dirac system where $v_{_{\rm NH}}=0$ from the outset~\cite{Isobe2012, RoyJHEP2016}. However, in such systems $x=v_{_{\rm H}}/c$, as $v_{_{\rm H}}=v_{_{\rm F}}$ therein.

The coupled RG flow equations from Eqs.~\eqref{eq:Flowc3D} and~\eqref{eq:FlowVF3D} support a single stable fixed point at $x=1$ or equivalently at $v_{_{\rm F}}=c$. However, in $d=3$, while the effective Fermi velocity for NH Dirac fermions increases logarithmically under coarse grain, the speed of light in Dirac crystals decreases logarithmically as we approach the deep infrared regime. As a result, even though the system ultimately always features a unique terminal velocity for all the participating degrees of freedom, thereby yielding a robust spacetime NH Lorentz symmetry (with both $v_{_{\rm H}}$ and $v_{_{\rm NH}}$ being nonzero), its magnitude is \emph{nonuniversal} and depends on the bare values of $v_{_{\rm H}}$ and $v_{_{\rm NH}}$. These outcomes are depicted in Fig.~\ref{fig:RGflow3D}. At the same time, we also note that the fine structure constant ($\alpha^{\rm 3D}_{_{\rm F}}$) decreases monotonically (but logarithmically slowly) as we approach the infrared regime, following Eq.~\eqref{eq:Flowalpha3D}.

\section{Summary and discussions}~\label{sec:summary}

To summarize, we develop an effective description of QED for NH Dirac materials, when the underlying fermoinic degrees of freedom, dispersing linearly with an effective Fermi velocity ($v_{_{\rm F}}$), interact with photons. Even though the collection of NH Dirac fermions is described by a Lorentz invariant NH operator [Eq.~\eqref{eq:NHDirac}], in the presence of fluctuating electromagnetic radiation the system generically lacks the Lorentz symmetry as $v_{_{\rm F}}$ and the speed of light $c$ are not equal at the ultraviolet or lattice scale. Nevertheless, we show that when such an interacting system approaches the deep infrared regime, quantum corrections [Fig.~\ref{fig:Self-energy}] always restore the spacetime Lorentz symmetry therein, where all the participating degrees of freedom possess a unique terminal velocity. See Figs~\ref{fig:RGflow2D} and~\ref{fig:RGflow3D}. As a two-dimensional Dirac system floats like a `brane' in a three-dimensional world, hosting photons, the speed of light does not renormalize due to the absence of any fermion mediated dynamic screening in $d=2$. Thus the terminal velocity in $d=2$ is always the speed of light in vacuum. By contrast, in $d=3$ both $v_{_{\rm F}}$ and $c$ flow under coarse grain and ultimately they reach a common, but \emph{nonuniversal} terminal velocity.

We realize that the renormalized velocity components of NH Dirac fermions, namely $v_{_{\rm H}}$ and $v_{_{\rm NH}}$, related to its effective Fermi velocity according to $v_{_{\rm F}}=\sqrt{v^2_{_{\rm H}}-v^2_{_{\rm NH}}}$, can become larger than the speed of light, while always maintaining $v_{_{\rm F}} \leq c$. This observation tempted us to consider a situation when the bare values of $v_{_{\rm H}}$ and $v_{_{\rm NH}}$ are larger than $c$, while that for $v_{_{\rm F}}<c$. In all these circumstances, we find a robust and generic restoration of the spacetime Lorentz symmetry. This observation may therefore suggest a possible generic emergence of the Lorentz symmetry in the NH universe of Dirac fermions, encompassing system-to-environment interactions.

The RG flow of the effective Fermi velocity of NH Dirac fermions leaves its signature on thermodynamic and transport observables. For example, the density of states and the specific heat in a $d$-dimensional NH Dirac system scale as $\varrho(E) \sim |E|^{d-1}/v^d_{_{\rm F}}$ and $C_v \sim T^{d}/v^d_{_{\rm F}}$, respectively. Here $T$ is the temperature. Therefore, increasing $v_{_{\rm F}}$ under coarse grain should manifest via decreasing magnitudes of these two thermodynamic quantities, while preserving their leading power-law scaling, resulting from the dynamic scaling exponent, measuring the relative scaling between energy and momentum, being locked at $z=1$. The frequency ($\omega$) dependent zero temperature longitudinal optical conductivity in two and three dimensional NH Dirac systems in units of $e^2/h$ are respectively $\sigma^{\rm 2D}_{\rm NH}=N_f \pi/4$ and $\sigma^{\rm 3D}_{\rm NH}=N_f \omega/(6 v_{_{\rm F}})$, where $h$ is the Planck's constant~\cite{RoyJuricicNHDSM2023}. So, $\sigma^{\rm 2D}_{\rm NH}$ remains unaffected by the renormalization of $v_{_{\rm F}}$. But, $\sigma^{\rm 3D}_{\rm NH}$ decreases at lower frequency while maintaining its hallmark $\omega$-linear scaling. Similarly, the optical shear viscosity in two- and three-dimensional NH Dirac systems scale as $\eta^{\rm 2D}_{\rm NH}=(N_f/128) \left( \omega/v_{_{\rm F}} \right)^2$ and $\eta^{\rm 3D}_{\rm NH}=(N_f/320\pi) \left( \omega/v_{_{\rm F}} \right)^3$ (for $\hbar=1$), respectively, showing similar power-law scaling as in Hermitian Dirac systems except $v_{_{\rm F}} \to v_{_{\rm H}}$ therein~\cite{RoyJuricicNHDSM2023, MooreViscosity2020}. Hence, shear viscosity decreases in two- and three-dimensional NH Dirac systems at lower frequency. In experiments, $T$ and $\omega$ act as infrared cutoffs for the RG flow of the velocity parameters.

In crystalline materials, the non-relativistic (NR) limit of our findings can be experimentally more pertinent, as typically $v_{_{\rm H}}$, $v_{_{\rm NH}}$, $v_{_{\rm F}} \ll c$ at the bare level, and the effect of the retardation and the restoration of the Lorentz symmetry develops logarithmically slowly. In the NR limit, the electron-photon interaction vertex is composed of only the instantaneous Coulomb part (density-density interaction) and is given by $-ie \gamma_\rho$ with $\gamma_\rho=(1, {\boldsymbol 0})$. The soft photon propagator in $d$ dimensions takes the form $D_{\rho \lambda}(i \omega, \vec{k},d)=\delta_{\rho \lambda}/|\vec{k}|^{d-1}$. The pertutbative corrections are obtained by taking the limit $x=v_{_{\rm F}}/c \to 0$ therein. Then in $d=2$, the RG flow equation reads as 
\begin{equation}
\beta_Q=\frac{\pi}{2} \: \alpha^{\rm 2D}_{\rm F, NR} \: Q
\end{equation}    
for $Q=v_{_{\rm H}}$ and $v_{_{\rm NH}}$, where $\alpha^{\rm 2D}_{\rm F, NR}=e^2/(16 \pi \varepsilon v_{_{\rm F}})$ is the fine structure constant for two-dimensional NR systems. Therefore, both the velocity components and the effective Fermi velocity increase logarithmically without any upper bound. Hence, the retardation effects in the form of the current-current interaction must be included in the formalism from the outset to showcase the emergence of a Lorentz symmetry. We arrive at similar conclusions in $d=3$, which follow from the RG flow equations 
\begin{equation}
\beta_{\alpha^{\rm 3D}_{\rm F, NR}}=-\frac{4}{3} N_f \; \left( \alpha^{\rm 3D}_{\rm F, NR} \right)^2 
\: \text{and} \:
\beta_Q=\frac{4}{3} \: \alpha^{\rm 3D}_{\rm F, NR} \: Q,
\end{equation} 
for $Q=v_{_{\rm H}}$ and $v_{_{\rm NH}}$. The fine structure constant for three-dimensional NR systems is $\alpha^{\rm 3D}_{\rm F, NR}=e^2/(8 \pi^2 \varepsilon v_{_{\rm F}})$. These outcomes closely mimic the ones in Hermitian Dirac systems, however, in terms of only $v_{_{\rm H}}$~\cite{Gonzalez1994, GoswamiOC2011, RoyJuricicOC2017}.

As far as the material realization of NH Dirac semimetals is concerned, designer electronic systems constitute the most promising ground, where a prominent two-dimensional Dirac system, graphene, has been engineered~\cite{Manoharan2012}. In graphene, conventional Dirac fermions stem from the nearest-neighbor hopping. Graphene accommodates a plethora of mass orders, represented by distinct mass matrices $M$~\cite{Ryu2009, Szabo2021}. The simplest of them is the charge-density-wave, featuring a density imbalance between two sublattices of the underlying honeycomb lattice~\cite{Semenoff1984}. Then the anti-Hermitian component of the NH Dirac operator in Eq.~\eqref{eq:NHDirac} also corresponds to nearest-neighbor hopping, however, with different amplitudes in the opposite directions, yielding non-Hermiticity in the system, as shown in Appendix~\ref{SMSec:NHgraphene}. Such an hopping imbalance can in principle be generated with two copies of optical honeycomb lattices~\cite{Esslinger2013}, occupied by neutral atoms living in the ground state and first excited state, which are coupled by running waves along three nearest-neighbor bond directions and the sites constituted by the excited state atoms undergoes a rapid loss. When the wavelength of the running wave is equal to the lattice spacing, a NH Dirac operator on optical honeycomb lattice with hopping imbalance between $A \rightarrow B$ and $B \rightarrow A$ directions can be realized. Here $A$ and $B$ denote two triangular sublattices of the honeycomb lattice. This proposal is a generalization of a similar one applied on a one-dimensional chain with left-right hopping imbalance, shown in Appendix~F of Ref.~\cite{GongPRX2018}. Similar engineering can also be executed in three-dimensional Dirac and Weyl materials, given the overall simplicity of the construction of the NH Dirac operator~\cite{RoyJuricicNHDSM2023} and the existence of mass matrices therein~\cite{SzaboJHEP2021}. However, for experimental verification of the theoretical predictions associated with the QED of NH Dirac fermions, the aforementioned proposal of NH Dirac materials needs to be implemented in electronic systems for which designer electronic platform is most suited, where conventional or Hermitian graphene has already been realized~\cite{Manoharan2012}. The present discussion should stimulate future experiments in this direction, unfolding novel QED and emergent NH Lorentz symmetry in NH Dirac materials.

\acknowledgments
This work was supported by NSF CAREER Grant No.\ DMR- 2238679 of B.R.\ and Dr.\ Hyo Sang Lee Graduate Fellowship from Lehigh University (S.A.M.). We are thankful to Vladimir Juri\v{c}i\'c for useful discussions and critical reading of the manuscript.


\appendix

\section{Fermionic self-energy}~\label{SMSec:FermionicSE}

The fermionic self-energy correction in a $d$-dimensional NH Diac material [see Fig.~\ref{fig:Self-energy}(a)] reads as 
\begin{equation}
\Sigma(i\nu,\vec{k})= (-i e)^2 \int \frac{d^D\vec{p}}{(2\pi)^D} \int^{\infty}_{-\infty} \frac{d\omega}{2 \pi} \:
\left[ \gamma_\rho G_{\rm F} (i \omega+i \nu, \vec{p}+\vec{k}) \gamma_\lambda D_{\rho \lambda}(i \omega, \vec{p},d) \right]. 
\end{equation}
The fermionic Green's function $G_{\rm F} (i \omega, \vec{p})$ and the propagator of the photon field $D_{\rho \lambda}(i \omega, \vec{p},d)$ are respectively shown in Eqs.~\eqref{eq:FemrionGreen} and~\eqref{eq:PhotonGreen}. The electron-photon vertex $\gamma_\rho$ is shown in Eq.~\eqref{eq:QEDvertex}. Then we find 
\begin{equation}
\Sigma(i\nu,\vec{k})=- \frac{e^2}{\varepsilon} \int \frac{d^D\vec{p}}{(2\pi)^D} \int^{\infty}_{-\infty} \frac{d\omega}{2 \pi} \:
\gamma_\rho \; \frac{i (\omega + \nu)-H_{\rm NH}(\vec{k}+\vec{p})}{\omega^2 + v^2_{_{\rm F}} k^2} \; \gamma_{\lambda} \; \frac{\delta_{\rho \lambda}}{\left[ \omega^2 + c^2 p^2 \right]^{\frac{d-1}{2}}}, 
\end{equation}
where $D=d-\epsilon$. The NH Dirac operator $H_{\rm NH}(\vec{k})$ is shown in Eq.~\eqref{eq:NHDirac}. After some straightforward algebra we find 
\begin{equation}
\gamma_\rho \left[ i (\omega + \nu)-H_{\rm NH}(\vec{k}+\vec{p}) \right] \gamma_{\lambda}\delta_{\rho \lambda}
=i (\omega + \nu) \left( 1-D \frac{v^2_{_{\rm F}}}{c^2}\right) - H_{\rm NH}(\vec{k}+\vec{p}) \left( 1+(D-2) \frac{v^2_{_{\rm F}}}{c^2} \right).
\end{equation}
We will shortly find out that the self-energy diagram yields logarithmic divergence $\sim 1/\epsilon$. Therefore, to capture the leading divergence, we can safely set $D=d$ in the above expression. Due to distinct structure of the photon propagator in $d=2$ and $d=3$, we display the computation of the fermionic self-energy for these two cases separately. 
\\

In two dimensions ($d=2$), after introducing the Feynman parameter ($y$) the self-energy contribution reads as 
\begin{eqnarray}
\Sigma(i\nu,\vec{k}) &=& -\frac{e^2}{2 \varepsilon} \int^1_0 dy \: (1-y)^{-1/2} \; \int \frac{d^D\vec{p}}{(2\pi)^D} \int^{\infty}_{-\infty} \frac{d\omega}{2 \pi} \left[ i (\omega+\nu)(1-2v^2_{_{\rm F}}/c^2) - H(\vec{k}+\vec{p}) \right] \nonumber \\
&\times& \left[ (\omega+y \nu)^2 +[c^2(1-y)+v^2_{_{\rm F}} y] \left(\vec{p} + \frac{y v^2_{_{\rm F}}}{c^2(1-y)+v^2_{_{\rm F}} y} \vec{k} \right)^2 + \Delta_1 \right]^{-3/2},
\end{eqnarray}
where 
\begin{equation}
\Delta_1=\frac{y (1-y) v^2_{_{\rm F}} c^2 k^2}{c^2(1-y)+v^2_{_{\rm F}} y} + y(1-y)\nu^2. 
\end{equation}
Next we make the following shift of variables 
\begin{equation}
\omega+y \nu \to \omega \:\:\:
\text{and} \:\:\:
\vec{p} + \frac{y v^2_{_{\rm F}}}{c^2(1-y)+v^2_{_{\rm F}} y} \vec{k} \to \vec{p}.
\end{equation}
Then after completing the integrations over the Matsubara frequency ($\omega$) and internal momentum ($\vec{p}$), we find 
\begin{eqnarray}
\Sigma(i\nu,\vec{k}) = -\frac{e^2}{8\pi^2 \varepsilon c^2} \Gamma\left( \frac{\epsilon}{2}\right) \int^1_0 dy \bigg[ 
i \nu \frac{1- 2 x^2}{1-y +y \; x^2} 
- \frac{H_{\rm NH}(\vec{k})}{ \left[1-y +y \; x^2 \right]^{-2}} \bigg] (1-y)^{1/2},
\end{eqnarray}
for $D=2-\epsilon$, where we have introduced a parameter $x=v_{_{\rm F}}/c$. Finally, we compute the integral over the Feynman parameter ($y$), leading to 
\begin{equation}
\Sigma(i\nu,\vec{k})=-\frac{e^2}{4\pi^2 \varepsilon c^2} \left[(i\nu) 2 \frac{1-2x^2}{1-x^2} \left\{ 1-\frac{x \cos^{-1}(x)}{\sqrt{1-x^2}}\right\} +  \frac{H(\vec{k})}{1-x^2} \left\{1- \frac{\cos^{-1}(x)}{x \sqrt{1-x^2}}\right\} \right] \frac{1}{\epsilon} + {\mathcal O}(1).
\end{equation}
From the final expression of the self-energy, we obtain the fermion field renormalization condition $Z_\Psi$, and the renormalization condition of two velocity parameters of NH Dirac fermions $Z_Q$, where $Q=v_{_{\rm H}}$ and $v_{_{\rm NH}}$ in two dimensions. 
\\

After introducing the Feynman parameter ($y$), the contribution from the fermionic self-energy diagram in $d=3$ reads as
\begin{eqnarray}
\Sigma(i\nu,\vec{k})&=&-\frac{e^2}{\varepsilon} \int^1_0 dy \; \int \frac{d^D\vec{p}}{(2\pi)^D} \int^{\infty}_{-\infty} \frac{d\omega}{2 \pi} \left[ i (\omega+\nu)(1-3 v^2_{_{\rm F}}/c^2) - H(\vec{k}+\vec{p})(1+ v^2_{_{\rm F}}/c^2) \right] \nonumber \\
&\times& \bigg[ (\omega+y \nu)^2 +[c^2(1-y)+v^2_{_{\rm F}} y] \left(\vec{p} + \frac{y v^2_{_{\rm F}}}{c^2(1-y)+v^2_{_{\rm F}} y} \vec{k} \right)^2 + \Delta_1 \bigg]^{-2}.
\end{eqnarray}
After completing the straightforward integrals over the Matsubara frequency ($\omega$) and the internal momentum ($\vec{p}$), in $D=3-\epsilon$ we find 
\begin{align}
\Sigma(i\nu,\vec{k})=-\frac{e^2 \; \Gamma\left(\epsilon/2\right)}{16 \pi^2 c^3 \varepsilon} \int^1_0 dy 
\left[(i \nu) \frac{(1-y) (1-3 x^2)}{[1-y+y \; x^2]^{3/2}}-H_{\rm NH}(\vec{k}) \frac{(1-y)(1+x^2)}{[1-y+y \; x^2]^{5/2}} \right],
\end{align} 
where $x=v_{_{\rm F}}/c$. Performing the integration over the Feynman parameter ($y$), we obtain the final expression for the self-energy correction in three dimensions, given by
\begin{eqnarray}
\Sigma(i\nu,\vec{k})=-\frac{e^2}{4 \pi^2 \varepsilon c^3} \left[ (i\nu) \; \frac{1-3 x^2}{(1+x)^2} - H_{\rm NH}(\vec{k}) \; \frac{(2+x)(1+x^2)}{3 x(1+x)^2} \right] \frac{1}{\epsilon} + {\mathcal O}(1). 
\end{eqnarray}
From this expression, we obtain the fermionic field renormalization $Z_\Psi$, and the renormalization condition for two velocity parameters $Z_Q$ for $Q=v_{_{\rm H}}$ and $v_{_{\rm NH}}$ in three dimensions.


\section{Self-energy for gauge fields}~\label{SMSec:PhotonSE}

The photon polarization bubble [see Fig.~\ref{fig:Self-energy}(b)] in $D$ dimensions reads as 
\begin{eqnarray}
\Pi_{\rho \lambda}(i\nu,\vec{k})=-(-ie)^2 \int \frac{d^D\vec{p}}{(2\pi)^D} \int^{\infty}_{-\infty} \frac{d\omega}{2 \pi} \: {\rm Tr}
\left[\gamma_\rho G_{\rm F} (i\omega,\vec{p}) \gamma_\lambda G_{\rm F} (i\omega+i\nu,\vec{p}+\vec{k}) \right].
\end{eqnarray}
We compute various components of the polarization bubble separately. For $\rho=\lambda=0$, we find 
\begin{eqnarray}
\Pi_{00}(i\nu,\vec{k})= 4 N_f e^2 \int \frac{d^D\vec{p}}{(2\pi)^D} \int^{\infty}_{-\infty} \frac{d\omega}{2 \pi} \:
\frac{-\omega(\omega+\nu)+v^2_{_{\rm F}} \vec{p}\cdot (\vec{k}+\vec{p})}{(\omega^2 + v^2_{_{\rm F}} p^2) (\omega^2 + v^2_{_{\rm F}} [\vec{p}+\vec{k}]^2)}.
\end{eqnarray}
After taking $v_{_{\rm F}}\vec{p} \to \vec{p}$ and $v_{_{\rm F}}\vec{k} \to \vec{k}$, and introducing the Feynman parameter ($y$), we obtain 
\begin{align}
\Pi_{00}(i\nu,\vec{k})=- \frac{e^2}{v^D_{_{\rm F}}} \int^1_0 dy \int \frac{d^D\vec{p}}{(2\pi)^D} \int^{\infty}_{-\infty} \frac{d\omega}{2 \pi} \:
\frac{4 N_f [\omega(\omega+\nu)-\vec{p}\cdot (\vec{k}+\vec{p})]}{\left[ (\omega+ y \nu)^2 + (\vec{p}+y \vec{k})^2 + y(1-y) (k^2+\nu^2)\right]^2}.
\end{align}
Next we perform a shift of variables according to $\omega+y \nu \to \omega$ and $\vec{p}+y \vec{k} \to \vec{p}$. Subsequently, the integrations over the Matsubara frequency ($\omega$) and internal momentum ($\vec{p}$) can be computed straightforwardly, leading to 
\begin{align}
\Pi_{00}(i\nu,\vec{k}) &=-\frac{2 N_f}{(4\pi)^{D/2}} \frac{e^2}{v^D_{_{\rm F}}} \int^1_0 dy  \: \frac{\Gamma(\frac{3-D}{2})}{\Gamma(3/2)} \: \int^1_0 dy y(1-y) \left[ y(1-y) (\nu^2+k^2) \right]^{\frac{D-3}{2}}.
\end{align}
The last expression yields a finite contribution for $D=2-\epsilon$, while producing a logarithmic or $1/\epsilon$ divergence when $D=3-\epsilon$, given by (after restoring the factor of $v_{_{\rm F}}$ via $k \to v_{_{\rm F}} k$)
\begin{equation}
\Pi_{00}(i\nu,\vec{k})=-\frac{4}{3} \frac{N_f}{v_{_{\rm F}}} \left( \frac{e^2}{8 \pi^2}\right) k^2 \frac{1}{\epsilon} + {\mathcal O} (1).
\end{equation}
Following the same procedure, next we compute $\Pi_{l0}(i\nu,\vec{k})$, where $l=1, \cdots, D$, given by 
\begin{align}
&\Pi_{l0}(i\nu,\vec{k}) = 4 N_f \frac{e^2}{c v^D_{_{\rm F}}} \int^1_0 dy \; [2y (1-y)] \int \frac{d^D\vec{p}}{(2\pi)^D} \int^{\infty}_{-\infty} \frac{d\omega}{2 \pi} \: \frac{\nu k_l}{\left(\omega^2 +p^2 + y(1-y)[\nu^2+k^2]\right)^2} \nonumber \\
&= N_f \frac{e^2}{c v^D_{_{\rm F}}} \left[ \frac{1}{(4\pi)^{D/2}} \frac{\Gamma(3/2-D/2)}{\Gamma(3/2)} \right] (\nu k_l) 
\int^1_0 dy \: \frac{2 y(1-y)}{\left[ y(1-y) (\nu^2+k^2) \right]^{3/2-D/2}}.
\end{align}
The last expression is finite near two spatial dimensions when we set $D=2-\epsilon$, while it shows a logarithmic or $1/\epsilon$ divergence for $D=3-\epsilon$, explicitly given by (upon restoring the factor of $v_{_{\rm F}}$ via $k \to v_{_{\rm F}} k$)
\begin{equation}
\Pi_{l0}(i\nu,\vec{k})=\frac{4}{3} \frac{N_f}{c v^2_{_{\rm F}}} \left( \frac{e^2}{8 \pi^2}\right) \; (\nu k_l) \; \frac{1}{\epsilon} + {\mathcal O} (1).
\end{equation}
Finally, we compute $\Pi_{lm}(i\nu,\vec{k})$, where $l,m=1,\cdots, D$. After completing the trace algebra, and shifting the integral variables to $\omega+y \nu \to \omega$ and $\vec{p}+y \vec{k} \to \vec{p}$ in terms of the Feynman parameter $y$, we find
\begin{eqnarray}
&&\Pi_{lm}(i\nu,\vec{k}) =-4 N_f \frac{e^2}{c^2} v^2_{_{\rm F}} \int^1_0 dy \int \frac{d^D\vec{p}}{(2\pi)^D} \int^{\infty}_{-\infty} \frac{d\omega}{2 \pi} \nonumber \\
&\times& \frac{\left[-\omega^2 + y(1-y) \nu^2 \right] \delta_{lm} + v^2_{_{\rm F}}\left[p_i p_j -y(1-y)k_i k_j \right] \left[ \delta_{li} \delta_{mj}- \delta_{lm} \delta_{ij} + \delta_{lj} \delta_{mi} \right]}{\left[ \omega^2 + v^2_{_{\rm F}} p^2 + v^2_{_{\rm F}} y(1-y) \left( k^2 + \nu^2/v^2_{_{\rm F}} \right) \right]^{2}}.
\end{eqnarray} 
The integrals over the Matsubara frequency ($\omega$) and internal momentum ($\vec{p}$) can now be computed straightforwardly. Upon performing these two integrals, we obtain 
\begin{eqnarray}
\Pi_{lm}(i\nu,\vec{k}) &=& -N_f \frac{e^2v_{_{\rm F}}}{c^2 (4\pi)^{D/2} \Gamma[3/2]} \int^1_0 dy \; \bigg\{ (1-D) \Gamma\left( \frac{1-D}{2}\right) (\Delta_2)^{\frac{D-1}{2}} \delta_{lk} \nonumber \\
&-& 2 y(y-1) k_l k_m \Gamma\left( \frac{3-D}{2} \right) (\Delta_2)^{\frac{D-3}{2}} \bigg\}, 
\end{eqnarray}
where $\Delta_2=y(1-y) [k^2 + \nu^2/v^2_{_{\rm F}}]$. Near two dimensions, $\Pi_{lm}(i\nu,\vec{k})$ is finite, whereas near three dimensions it shows logarithmic or $1/\epsilon$ divergence, captured by 
\begin{eqnarray}
\Pi_{lm}(i\nu,\vec{k})=-\frac{4}{3} N_f \frac{e^2 v_{_{\rm F}}}{8 \pi^2 c^2} \: \left[ \left( k^2 + \frac{\nu^2}{v^2_{_{\rm F}}} \right) \delta_{lm} - k_l k_m \right] \; \frac{1}{\epsilon} + {\mathcal O}(1).
\end{eqnarray}

Upon collecting all the components of the polarization tensor, it can be compactly written as 
\begin{equation}
\Pi_{\rho \lambda}(i\nu,\vec{k})= - \frac{4}{3} N_f \left( \frac{e^2}{8 \pi^2} \right) \left[ k^2 \delta_{\rho \lambda} -k_\rho k_\lambda \right] \left( \frac{v_{_{\rm F}}}{c} \right)^{2-\delta_{\rho,0}-\delta_{\lambda,0}} \frac{k^{\epsilon}}{v^3_{_{\rm F}}} \; \frac{1}{\epsilon} + {\mathcal O} (1), 
\end{equation}
where $k_\rho=(k_0, v_{_{F}} \vec{k})=(i\nu,v_{_{F}} \vec{k})$. Performing a Wick rotation we can go to the real time from the imaginary time, yielding the renormalized photon polarization 
\begin{equation}
\Pi^{\rm R}_{\rho \lambda}=\frac{1}{2} A_\rho \left[ \frac{4}{3} N_f \left( \frac{e^2}{8 \pi^2} \right) \left[ k^2 {\mathcal G}_{\rho \lambda} -k_\rho k_\lambda \right] \left( \frac{v_{_{\rm F}}}{c} \right)^{2-\delta_{\rho,0}-\delta_{\lambda,0}} \frac{k^{\epsilon}}{v^3_{_{\rm F}}} \; \frac{1}{\epsilon} \right] A_\lambda + {\mathcal O}(1),
\end{equation}
where ${\mathcal G}_{\rho \lambda}={\rm Dig}.(1,-1,-1,-1)$ is the metric tensor. Then the renormalized Maxwellian Lagrangian takes the form 
\begin{align}
L^{\rm R}_{\rm EM}&=\varepsilon {\bf E}^2-\frac{{\bf B}^2}{\mu}
= \varepsilon \left[-A^2 \frac{k^2_0}{c^2} -A^2_0 k^2 - 2 A_0 A_j \frac{k_0}{c} k_j \right] \: \left(1-\frac{4}{3} N_f \frac{e^2}{8 \pi^2 \varepsilon c^2 v_{_{\rm F}}} \; \frac{1}{\epsilon} \right) \nonumber \\
&- \frac{1}{\mu c^2} \left[ A^2_1(k^2_2+k^2_3) + A^2_2(k^2_3+k^2_1) + A^2_3(k^2_3+k^2_1) - 2 A_i A_j k_i k_j \right] \: \left(1-\frac{4}{3} N_f \frac{e^2 \mu v_{_{\rm F}}}{8 \pi^2 c^2} \; \frac{1}{\epsilon} \right),
\end{align}
which follows from the definition of the electric (${\bf E}$) and magnetic (${\bf B}$) fields in terms of the vector potentials, namely 
\begin{equation}
{\bf E}=-\frac{1}{c} \frac{\partial \vec{A}}{\partial t} -{\boldsymbol \nabla} A_0 
\quad \text{and} \quad 
{\bf B}=\frac{1}{c} {\boldsymbol \nabla} \times \vec{A}, 
\end{equation}
respectively. From the expression of $L^{\rm R}_{\rm EM}$, we obtain the renormalization conditions for the permittivity ($\varepsilon$) and permeability ($\mu$) of the medium [see Eq.~\eqref{eq:RGepsilonmu}], which then leads to the renormalization condition of the speed of light ($c$) in the medium, and its renormalization group flow equation [see Eq.~\eqref{eq:Flowc3D}].


\section{Vertex correction and Ward identity}~\label{SMSec:Vertex}

The correction to the electron-photon vertex [see Fig.~\ref{fig:Self-energy}(c)] in $D$ dimensions reads as
\begin{equation}
\Gamma_\mu=(-ie)^3 \; \int \frac{d^D\vec{p}}{(2\pi)^D} \int^{\infty}_{-\infty} \frac{d\omega}{2 \pi}  \:
\gamma_\lambda G_{\rm F} (i \nu_{p} -i \omega, \vec{k}_{p}-\vec{p}) \gamma_\mu G_{\rm F} (i \nu_{p'} -i \omega, \vec{k}_{p'}-\vec{p}) D_{\rho \lambda} (i\omega, \vec{p},d), 
\end{equation}
where $\mu=0,\cdots,d$ and $D=d-\epsilon$. Here, we are interested in the divergent pieces of $\Gamma_\mu$ for which we can set $\nu_p=\nu_{p'}=0$ and $\vec{k}_p=\vec{k}_{p'}=0$ in the numerator. Then after some straightforward algebra the part of the numerator, contributing to the divergent part of $\Gamma_\mu$, denoted by $N^{\rm div}_\mu$, simplifies to 
\begin{align}
N^{\rm div}_\mu=\left( \left[ -\omega^2 + v^2_{_{\rm F}} p^2 \right] \left( 1-D \frac{v^2_{_{\rm F}}}{c^2} \right), 
-\frac{i}{c} \left( 1-(2-D) \frac{v^2_{_{\rm F}}}{c^2} \right) \left[ -\omega^2 + \frac{2-D}{D} v^2_{_{\rm F}} p^2 \right] \gamma_j \right),
\end{align} 
where $\gamma_j=v_{_{\rm H}} \Gamma_j + v_{_{\rm NH}} M \Gamma_j$ and $j=1,\cdots, d$. Next we cast the denominator of $\Gamma_\mu$ in a symmetric form by introducing the Feynman parameters ($w$, $y$ and $z$). We show these steps for $d=2$ and $d=3$ separately. In $d=2$, 
\begin{align}
&\frac{1}{(\nu_{p} - \omega)^2 + v^2_{_{\rm F}} (\vec{k}_{p}-\vec{p})^2} \;
\frac{1}{(\nu_{p'} - \omega)^2 + v^2_{_{\rm F}} (\vec{k}_{p'}-\vec{p})^2} \;
\frac{1}{\left[\omega^2 + c^2 p^2 \right]^{1/2}} 
= \frac{\Gamma(5/2)}{\Gamma(1/2)} \int^1_0 dw \int^1_0 dy \nonumber \\  
&\int^1_0 dz  \frac{\delta(w+y+z-1) \: z^{-1/2}}{\left\{ w \left[ (\nu_{p} - \omega)^2 + v^2_{_{\rm F}} (\vec{k}_{p}-\vec{p})^2\right] + y \left[ (\nu_{p'} - \omega)^2 + v^2_{_{\rm F}} (\vec{k}_{p'}-\vec{p})^2 \right]+ z \left[ \omega^2 + c^2 p^2 \right]\right\}^{5/2}} \nonumber \\
&= \frac{3}{4} \int^1_0 dz \int^{1-z}_0 dy \frac{z^{-1/2}}{\left[ \omega^2 + \left[z c^2 + (1-z)v^2_{_{\rm F}} \right] p^2 + \Delta_3 \right]^{5/2}},
\end{align}
after the following shift of variables 
\begin{equation}
\omega- (1-y-z) \nu_{p} - y \nu_{p'} \to \omega 
\quad \text{and} \quad
\vec{p}- \frac{(1-y-z) \vec{k}_{p} + y \vec{k}_{p'}}{z c^2 + (1-z)v^2_{_{\rm F}}} \to \vec{p}, \nonumber 
\end{equation}
where 
\begin{eqnarray}
\Delta_3 &=& (1-y-z) \nu^2_{p} + y \nu^2_{p'} -\left[ (1-y-z) \nu_{p} + y \nu_{p'} \right]^2 \nonumber \\
&+& v^2_{_{\rm F}} \left[(1-y-z) \vec{k}^2_{p} + y \vec{k}^2_{p'} \right] -v^4_{_{\rm F}} \frac{[(1-y-z) \vec{k}_{p} + y \vec{k}_{p'}]^2}{z c^2 + (1-z)v^2_{_{\rm F}}}.
\end{eqnarray}
The above mentioned shift of variables does not affect $N^{\rm div}_\mu$, which is now defined in terms of the shifted variables $\omega$ and $\vec{p}$. Then a straightforward calculation leads to the divergence components of $\Gamma_\mu$, denoted by $\Gamma^{\rm div}_\mu$, given by 
\begin{align}
\Gamma^{\rm div}_\mu = (-ie) \frac{e^2}{4\pi^2 \varepsilon c^2}  \bigg(- \frac{2(1-2 x^2)}{1-x^2} 
\left\{ 1- \frac{x \cos^{-1}(x)}{\sqrt{1-x^2}}\right\}, -\frac{i}{c} \gamma_j \; \frac{1}{1-x^2} \left\{ 1- \frac{\cos^{-1}(x)}{x\sqrt{1-x^2}}\right\} \bigg) \; \frac{1}{\epsilon}.
\end{align} 
In $d=3$, following the identical procedure we arrive at the following expression for the denominator 
\begin{eqnarray}
&&\frac{1}{(\nu_{p} - \omega)^2 + v^2_{_{\rm F}} (\vec{k}_{p}-\vec{p})^2} \:
\frac{1}{(\nu_{p'} - \omega)^2 + v^2_{_{\rm F}} (\vec{k}_{p'}-\vec{p})^2} \:
\frac{1}{\omega^2 + c^2 p^2} 
= 2! \int^1_0 dw \int^1_0 dy \nonumber \\
&\times& \int^1_0 dz \; \frac{\delta(w+y+z-1)}{\left\{ w \left[ (\nu_{p} - \omega)^2 + v^2_{_{\rm F}} (\vec{k}_{p}-\vec{p})^2\right] + y \left[ (\nu_{p'} - \omega)^2 + v^2_{_{\rm F}} (\vec{k}_{p'}-\vec{p})^2 \right]+ z \left[ \omega^2 + c^2 p^2 \right]\right\}^{3}} \nonumber \\
&=& 2! \int^1_0 dz \int^{1-z}_0 dy \frac{z^{-1/2}}{\left[ \omega^2 + \left[z c^2 + (1-z)v^2_{_{\rm F}} \right] p^2 + \Delta_3 \right]^{3}},
\end{eqnarray}
in terms of the shifted variables $\omega$ and $\vec{p}$. Then we arrive at the following expression for the divergent piece of the vertex function
\begin{eqnarray}
\Gamma^{\rm div}_\mu=(-ie) \frac{e^2}{4\pi^2 \varepsilon c^3} \left(- \frac{1-3 x^2}{(1+x)^2}, -\frac{i}{c} \gamma_j \; \frac{(2+x)(1+x^2)}{3x(1+x)^2} \right) \; \frac{1}{\epsilon}.
\end{eqnarray} 
In both two and three dimensions, we find that 
\begin{equation}
\frac{\partial \Sigma(i\nu,\vec{k})}{\partial(i\nu)}=(-i e)^{-1} \Gamma^{\rm div}_0 
\quad \text{and} \quad
\frac{\partial \Sigma(i\nu,\vec{k})}{\partial(k_j)}=(-i e)^{-1} \left(-\frac{i}{c}\right)^{-1} \Gamma^{\rm div}_j
\end{equation}
for $j=1, \cdots, d$, which is in agreement with the Ward identity~\cite{ward1950}. 
\\


\section{Non-Hermitian Dirac operator in graphene: An example}~\label{SMSec:NHgraphene}

In this Appendix, we outline the simplest possible realization of the NH Dirac operator on graphene's honeycomb lattice. The tight-binding Bloch Hamiltonian with only nearest-neighbor hopping amplitude ($t_0$) in this system takes the form  
\begin{equation}~\label{eq:TBHoneycomb}
h^{\rm lat}_0({\bf k})= t_0 \; \left( \begin{array}{cc}
0 & f({\bf k}) \\
f^\star({\bf k}) & 0
\end{array} \right), 
\end{equation}
written in the basis of a two component spinor $\Psi^\top_{\bf k}=(c_A, c_B)({\bf k})$, where $c_A({\bf k})$ and $c_B({\bf k})$ are the fermionic annihilation operators on the sites of two interpenetrating triangular sublattices $A$ and $B$, respectively, with momentum ${\vec{k}}$ and $f({\bf k})=\exp[i {\bf k} \cdot {\bf b}_1] + \exp[i {\bf k} \cdot {\bf b}_2] + \exp[i {\bf k} \cdot {\bf b}_3]$. Here, ${\bf b}_1=(1/\sqrt{3},1)a/2$, ${\bf b}_2=(1/\sqrt{3},-1)a/2$ and ${\bf b}_3=(-1/\sqrt{3},0)a$ are the nearest-neighbor vectors, $a$ is the lattice spacing and `$\star$' denotes the complex conjugation. 
\\

 The Dirac mass, resulting from the staggered density between two sublattices of the honeycomb lattice is given by $h^{\rm lat}_{M}={\rm diag.}(1,-1)$~\cite{Semenoff1984}. Notice that $\{h^{\rm lat}_0({\bf k}), h^{\rm lat}_{M}\}=0$ and thus $h^{\rm lat}_{M}$ acts as a mass for Dirac fermions. Therefore, in terms of the hopping and mass Hamiltonian, we can define the following NH operator [following Eq.~\eqref{eq:NHDirac}] 
\begin{equation}
h^{\rm lat}_{\rm NH}({\bf k})= h^{\rm lat}_0({\bf k}) + \alpha h^{\rm lat}_{M} h^{\rm lat}_0({\bf k})= t_0 \left( 
\begin{array}{cc}
0 & (1+\alpha) f({\bf k}) \\
(1-\alpha) f^\star({\bf k}) & 0 
\end{array} \right),
\end{equation}  
where the parameter $\alpha$ is real and $|\alpha|<1$. This NH operator therefore represents a hopping imbalance between the directions $A \to B$ and $B \to A$. Namely the hopping amplitude in the $A \to B$ direction is $t_0 (1+\alpha)$ and in the $B \to A$ direction is $t_0 (1-\alpha)$, which gives rise to the non-Hermiticity. For simplicity, we neglect the spin degrees of freedom, which only leads to a mare doubling of $h^{\rm lat}_{\rm NH}({\bf k})$. The NH operator $h^{\rm lat}_{\rm NH}({\bf k})$ conforms to the NH Dirac operator shown in Eq.~\eqref{eq:NHDirac}, which can be seen after expanding $f({\bf k})$ and $f^\star({\bf k})$ around two inequivalent valleys at $\pm {\bf K}$, where ${\bf K}=(1,1/\sqrt{3})(2\pi/\sqrt{3}a)$~\cite{Semenoff1984}, leading to 
\begin{eqnarray}
H_{\rm NH} &=& v_{_{\rm H}} \left[\gamma_{_{31}} k_x + \gamma_{_{02}} k_y \right] + v_{_{\rm NH}} \gamma_{_{03}} \left[\gamma_{_{31}} k_x + \gamma_{_{02}} k_y \right] 
\equiv \left[ v_{_{\rm H}} + v_{_{\rm NH}} M \right] \; \left[ \Gamma_1 k_x + \Gamma_2 k_y \right] \nonumber \\
&\equiv& v_{_{\rm H}}h_0 + v_{_{\rm NH}} M h_0,
\end{eqnarray}  
where $v_{_{\rm H}}=\sqrt{3} t_0 a/2$ and $v_{_{\rm NH}}=\alpha \sqrt{3} t_0 a/2$, such that $v_{_{\rm H}}>v_{_{\rm NH}}$ as long as $|\alpha|<1$. Various matrix operators are $h_0=\Gamma_1 k_x + \Gamma_2 k_y$, representing the standard Dirac Hamiltonian, $\Gamma_1=\gamma_{_{31}}$, $\Gamma_2=\gamma_{_{02}}$, $M=\gamma_{_{03}}$, where $\gamma_{_{\mu \nu}}=\tau_\mu \beta_\nu$. The set of Pauli matrices $\{ \tau_\mu\}$ and $\{ \beta_\mu \}$ operate on the valley and sublattice degrees of freedom, respectively. Therefore, hopping imbalance between the $A \to B$ and $B \to A$ directions on honeycomb lattice yields one of realizations of Lorentz invariant NH Dirac operator confirming to Eq.~\eqref{eq:NHDirac}.  \\

{\bf Open Access}.~This article is distributed under the terms of the Creative Commons Attribution License (CC-BY 4.0), which permits any use, distribution and reproduction in any medium, provided the original author(s) and source are credited.


\bibliographystyle{JHEP}

\bibliography{ReferencesNHQED}

\providecommand{\href}[2]{#2}\begingroup\raggedright\begin{thebibliography}{10}

\bibitem{Peskin2019}
M.~E. Peskin and D.~V. Schroeder, \emph{An Introduction to Quantum Field
  Theory}.
\newblock CRC Press, 2019.

\bibitem{Schwinger1948}
J.~Schwinger, \emph{{On Quantum-Electrodynamics and the Magnetic Moment of the
  Electron}}, \href{https://doi.org/10.1103/PhysRev.73.416}{\emph{Phys. Rev.}
  {\bfseries 73} (1948) 416}.

\bibitem{Nielsen1978}
H.~Nielsen and M.~Ninomiya, \emph{{$\beta$-Function in a non-covariant
  Yang-Mills theory}},
  \href{https://doi.org/https://doi.org/10.1016/0550-3213(78)90341-3}{\emph{Nucl.
  Phys. {\bf B}} {\bfseries 141} (1978) 153}.

\bibitem{Kostelecky2011}
V.~A. Kosteleck\'y and N.~Russell, \emph{{Data tables for Lorentz and $CPT$
  violation}}, \href{https://doi.org/10.1103/RevModPhys.83.11}{\emph{Rev. Mod.
  Phys.} {\bfseries 83} (2011) 11}.

\bibitem{Dirac1928}
P.~A.~M. Dirac, \emph{{The quantum theory of the electron}},
  \href{https://doi.org/10.1098/rspa.1928.0023}{\emph{Proc. R. Soc. Lond. A}
  {\bfseries 117} (1928) 610}.

\bibitem{Castroneto2009}
A.~H. Castro~Neto, F.~Guinea, N.~M.~R. Peres, K.~S. Novoselov and A.~K. Geim,
  \emph{{The electronic properties of graphene}},
  \href{https://doi.org/10.1103/RevModPhys.81.109}{\emph{Rev. Mod. Phys.}
  {\bfseries 81} (2009) 109}.

\bibitem{Balatsky2014}
T.~O. {Wehling}, A.~M. {Black-Schaffer} and A.~V. {Balatsky}, \emph{{Dirac
  materials}}, \href{https://doi.org/10.1080/00018732.2014.927109}{\emph{Adv.
  Phys.} {\bfseries 63} (2014) 1}.

\bibitem{Armitage2018}
N.~P. Armitage, E.~J. Mele and A.~Vishwanath, \emph{{Weyl and Dirac semimetals
  in three-dimensional solids}},
  \href{https://doi.org/10.1103/RevModPhys.90.015001}{\emph{Rev. Mod. Phys.}
  {\bfseries 90} (2018) 015001}.

\bibitem{Chadha1983}
S.~Chadha and H.~Nielsen, \emph{{Lorentz invariance as a low energy
  phenomenon}},
  \href{https://doi.org/https://doi.org/10.1016/0550-3213(83)90081-0}{\emph{Nucl.
  Phys. {\bf B}} {\bfseries 217} (1983) 125}.

\bibitem{Gonzalez1994}
J.~Gonz\'alez, F.~Guinea and M.~Vozmediano, \emph{{Non-Fermi liquid behavior of
  electrons in the half-filled honeycomb lattice (A renormalization group
  approach)}},
  \href{https://doi.org/https://doi.org/10.1016/0550-3213(94)90410-3}{\emph{Nucl.
  Phys. {\bf B}} {\bfseries 424} (1994) 595}.

\bibitem{Isobe2012}
H.~Isobe and N.~Nagaosa, \emph{{Theory of a quantum critical phenomenon in a
  topological insulator: (3+1)-dimensional quantum electrodynamics in solids}},
  \href{https://doi.org/10.1103/PhysRevB.86.165127}{\emph{Phys. Rev. B}
  {\bfseries 86} (2012) 165127}.

\bibitem{RoyJHEP2016}
B.~Roy, V.~Juri{\v{c}}i{\'{c}} and I.~F. Herbut, \emph{{Emergent Lorentz
  symmetry near fermionic quantum critical points in two and three
  dimensions}}, \href{https://doi.org/10.1007/JHEP04(2016)018}{\emph{J. High
  Energy Phys.} {\bfseries 2016} (2016) 18}.

\bibitem{Esslinger2013}
T.~Uehlinger, G.~Jotzu, M.~Messer, D.~Greif, W.~Hofstetter, U.~Bissbort et~al.,
  \emph{{Artificial Graphene with Tunable Interactions}},
  \href{https://doi.org/10.1103/PhysRevLett.111.185307}{\emph{Phys. Rev. Lett.}
  {\bfseries 111} (2013) 185307}.

\bibitem{GongPRX2018}
Z.~Gong, Y.~Ashida, K.~Kawabata, K.~Takasan, S.~Higashikawa and M.~Ueda,
  \emph{{Topological Phases of Non-Hermitian Systems}},
  \href{https://doi.org/10.1103/PhysRevX.8.031079}{\emph{Phys. Rev. X}
  {\bfseries 8} (2018) 031079}.

\bibitem{Manoharan2012}
K.~K. Gomes, W.~Mar, W.~Ko, F.~Guinea and H.~Manoharan, \emph{{Designer Dirac
  fermions and topological phases in molecular graphene}},
  \href{https://doi.org/10.1038/nature10941}{\emph{Nature (London)} {\bfseries
  483} (2012) 306}.

\bibitem{RoyJuricicNHDSM2023}
V.~{Juricic} and B.~{Roy}, \emph{{Yukawa-Lorentz Symmetry in Non-Hermitian
  Dirac Materials}}, {\emph{ArXiv e-prints: 2308.16907} (2023) },
  [\href{https://arxiv.org/abs/2308.16907}{{\ttfamily 2308.16907}}].

\bibitem{Bernard2002}
D.~Bernard and A.~LeClair, \emph{{A Classification of Non-Hermitian Random
  Matrices}}.
\newblock Springer Netherlands, Dordrecht, 2002.

\bibitem{Jackson1999}
J.~D. Jackson, \emph{Classical Electrodynamics}.
\newblock John Wiley \& Sons, 1999.

\bibitem{ward1950}
J.~C. Ward, \emph{{An Identity in Quantum Electrodynamics}},
  \href{https://doi.org/10.1103/PhysRev.78.182}{\emph{Phys. Rev.} {\bfseries
  78} (1950) 182}.

\bibitem{MooreViscosity2020}
M.~Moore, P.~Sur\'owka, V.~Juri\ifmmode \check{c}\else
  \v{c}\fi{}i\ifmmode~\acute{c}\else \'{c}\fi{} and B.~Roy, \emph{{Shear
  viscosity as a probe of nodal topology}},
  \href{https://doi.org/10.1103/PhysRevB.101.161111}{\emph{Phys. Rev. B}
  {\bfseries 101} (2020) 161111}.

\bibitem{GoswamiOC2011}
P.~Goswami and S.~Chakravarty, \emph{{Quantum Criticality between Topological
  and Band Insulators in $3+1$ Dimensions}},
  \href{https://doi.org/10.1103/PhysRevLett.107.196803}{\emph{Phys. Rev. Lett.}
  {\bfseries 107} (2011) 196803}.

\bibitem{RoyJuricicOC2017}
B.~Roy and V.~Juri\ifmmode \check{c}\else \v{c}\fi{}i\ifmmode~\acute{c}\else
  \'{c}\fi{}, \emph{{Optical conductivity of an interacting Weyl liquid in the
  collisionless regime}},
  \href{https://doi.org/10.1103/PhysRevB.96.155117}{\emph{Phys. Rev. B}
  {\bfseries 96} (2017) 155117}.

\bibitem{Ryu2009}
S.~Ryu, C.~Mudry, C.-Y. Hou and C.~Chamon, \emph{{Masses in graphenelike
  two-dimensional electronic systems: Topological defects in order parameters
  and their fractional exchange statistics}},
  \href{https://doi.org/10.1103/PhysRevB.80.205319}{\emph{Phys. Rev. B}
  {\bfseries 80} (2009) 205319}.

\bibitem{Szabo2021}
A.~L. Szab\'o and B.~Roy, \emph{{Extended Hubbard model in undoped and doped
  monolayer and bilayer graphene: Selection rules and organizing principle
  among competing orders}},
  \href{https://doi.org/10.1103/PhysRevB.103.205135}{\emph{Phys. Rev. B}
  {\bfseries 103} (2021) 205135}.

\bibitem{Semenoff1984}
G.~W. Semenoff, \emph{{Condensed-Matter Simulation of a Three-Dimensional
  Anomaly}}, \href{https://doi.org/10.1103/PhysRevLett.53.2449}{\emph{Phys.
  Rev. Lett.} {\bfseries 53} (1984) 2449}.

\bibitem{SzaboJHEP2021}
A.~L. Szab$\acute{\mbox{o}}$ and B.~Roy, \emph{{Emergent chiral symmetry in a
  three-dimensional interacting Dirac liquid}},
  \href{https://doi.org/10.1007/JHEP01(2021)004}{\emph{J. High Energy Phys.}
  {\bfseries 2021} (2021) 27}.

\end{thebibliography}\endgroup

\end{document}